\documentclass[onecolumn,preprintnumbers,amsmath,amssymb,aps,floatfix]{revtex4-1}
\usepackage[utf8]{inputenc}
\usepackage{amsmath,amsfonts,amssymb}
\usepackage{graphicx}
\usepackage{textcomp}
\usepackage[colorlinks]{hyperref}
\usepackage{setspace}
\usepackage{tocloft}
\usepackage[normalem]{ulem}
\usepackage{color,soul}

\begin{document}
\title{Label-free microscopy of mitotic chromosomes using the polarization Orthogonality Breaking technique}


\author{R. Desapogu,$^{1,2,\#}$}
\author{G. Le Marchand$^{1,\#}$}
\author{R. Smith,$^{1}$}
\author{P. Ray,$^{2}$}
\author{E. Gillier,$^{1}$}
\author{S. Dutertre,$^{3}$}
\author{M. Alouini,$^{2}$}
\author{M. Tramier,$^{1,3}$}
\author{S. Huet,$^{1,3,4,*}$}
\author{J. Fade,$^{2,*}$\\}
\email{julien.fade@univ-rennes1.fr, and sebastien.huet@univ-rennes1.fr\\ $^\#$ \it These authors contributed equally to this work}

 \affiliation{$^1$ Univ Rennes, CNRS, IGDR (Genetics and Development Institute of Rennes), UMR 6290, F-35000 Rennes, France}
 \affiliation{$^2$ Univ Rennes, CNRS, Institut FOTON - UMR 6082, F-35000 Rennes, France}
 \affiliation{$^3$ Univ Rennes, BIOSIT, UMS CNRS 3480, US INSERM 018, F-35000 Rennes, France}
 \affiliation{$^4$ Institut Universitaire de France}



\begin{abstract}

  The vast majority of the microscopy methods currently available to
  study biological samples require staining prior to
  imaging. Nevertheless, the ability to reveal specific cell
  structures or organelles in a label-free manner remains desirable in
  different contexts.  Polarization microscopy has long been
  considered as an interesting alternative to fluorescence-based
  methods in order to gain specificity on the imaged biological
  samples. In this work, we show how an original polarization imaging
  technique, implementing micro-wave photonics and referred to as
  orthogonality-breaking (OB) microscopy, can provide informative
  polarization images from a single scan of the cell sample in a fast
  and sensitive way. For OB imaging, the sample is probed with a laser
  setup  simultaneously generating two orthogonal polarizations shifted
  in frequency by a few tens of MHz. If the imaged samples display
  some polarimetric properties, the orthogonality between the two
  polarizations is broken, leading to a beatnote interference signal
  that can be detected with a fast detector. The comparison of the
  images of various cell lines at different cell-cycle stages obtained
  by OB polarization microscopy and fluorescence confocal images shows
  that an endogenous polarimetric contrast arizes on compacted
  chromosomes during cell division. This technique paves the way to
  label-free real-time polarization imaging of mitotic chromosomes
  with further potential applications in histology and cancer
  diagnosis.
\end{abstract}

\maketitle







\section*{Introduction}
Over the last decades, the relentless development of novel optical
imaging techniques addressing specific issues of the biology or
biomedical community has given rise to a huge number of unconventional
imaging techniques. Several of these were granted with worldwide
commercial and scientific success such as Optical Coherence Tomography,
advanced confocal fluorescence microscopy,
non-linear microscopy, phase imaging, etc. However, a number of interesting
challenges still remain open in this interdisciplinary research
field. For some of these challenges, polarimetry can inspire novel
solutions to overcome some of the remaining bottlenecks. For instance,
the study of cellular biological mechanisms is overwhelmingly
performed with confocal laser scanning fluorescence microscopy, which
allows the structures of interest to be imaged at high spatio-temporal
resolution. Nevertheless, as for any fluorescence-based method,
confocal microscopy requires the labeling of the sample using either
fluorescent compounds displaying specific localizations whithin the
cells, dyes coupled to antibodies or recombinant chimeric constructs
composed of a fluorescent protein fused to a protein of interest.
Although this labeling step has proven to be usable in multiple contexts,
it also suffers some limitations. First, \textit{per se}, it requires
an additional, potentially lenghty, step in the sample preparation,
thus limiting the throughput of the experiments. Second, due to the
crosstalk between the emission spectra of the fluorescent dyes, it is
often difficult to image simultaneously more than four different
structures within the same sample. Finally, in the context of the
observation of living samples, the labeling procedure may induce
cytotoxicity and interfere with the tracked biological
mechanisms. These different drawbacks justify the need for the
development of label-free microscopy techniques. Among the techniques
that are currently investigated (non-linear microscopy, quantitative
phase microscopy, diffraction tomography…), polarimetric approaches
can provide multidimensional physical information on the samples,
which could thus convey also some selectivity in order to identify
cell constituents \cite{old96,old98,ino08,mor08z,zhe19}, investigate
the internal organization of biological structures \cite{gho11,bra11,
  ell14, ala15} or discriminate unhealthy tissues
\cite{chu07,sal09,pie11, pie13} without the need for labeling. Most of
the time however, polarimetric imaging is not given preference, due to
the poor sensitivity of most systems and to the complex calibration
and lengthy acquisition procedures of standard polarimetric techniques
based on the sequential acquisition of several
images\cite{gho11,qi17,kuh01}.

In this context, a direct, sensitive and fast polarimetric imaging
technique, referred to as “orthogonality-breaking” (OB) polarimetric
sensing has been proposed a few years ago \cite{fad12,sch14,ort15}. It
is based on a microwave photonics approach, and allows different
polarimetric properties (birefringence, dichroism, and depolarization)
to be identified from a single sample scan using the appropriate
detection modality \cite{par20}. In this paper, we report how such a
technique has been implemented on a commercial fluorescence confocal
microscope set up. We demonstrate that this new imaging modality can
be used to observe mitotic chromosomes at different stages of the cell
division with high contrast and at the spatial resolution allowed by
the microscope setup, i.e., approximately $\simeq 150$~nm within the
focal plane. Being able to monitor such cellular structures can prove
useful to quickly identify proliferating cells within a biological
sample or to identify cells displaying altered architecture of their
genomic material, a feature that is shared by multiple tumor cells
\cite{zin04}.

\section*{Materials and methods}
\subsection*{Theory}\label{sec:theory}
\subsubsection*{Principle of orthogonality-breaking polarimetric sensing}

OB polarimetric imaging is based on the use of a specific
dual-frequency dual-polarization (DFDP) laser illumination to probe
the sample. The frequency difference imposed between the two
polarization components of equal intensities must lie in the
radiofrequency (RF) range, typically 10's to 100's of MHz, in order to
match the typical bandwidth of common photodetectors (PD)
(photodiodes, avalanche photodiodes (APD),...), and avoid chromatic
dispersion effects during light propagation in the setup and the
sample \cite{fad12}. It has been shown that the illumination
polarization states should preferably be left/right circular
\cite{ort15}, in which case the fast temporal evolution of the
electric field produced corresponds to a linear
polarization state rotating at RF frequency. Due to the imposed orthogonality
between the two components oscillating at distinct frequencies, the
intensity of such beam has a constant value, even when measured with a
fast PD.

When such a beam  interacts with a sample, the intensity
detected upon interaction with it remains constant, unless some
polarization ``orthogonality-breaking'' takes place. In this
configuration initially proposed in \cite{fad12}, such OB only occurs
when light interacts with a dichroic sample (absorption anisotropy):
the detected intensity shows a fast RF temporal modulation (beatnote),
due to the interference of the two frequency components of the probe
beam \cite{ort15,ort16}. Interestingly, we showed previously that the
normalized amplitude of the beatnote (ratio of the beatnote amplitude
(AC) by the average intensity value (DC)), which we shall refer to in
the following as orthogonality-breaking contrast (OBC) is a direct
measure of the diattenuation rate of the dichroic sample
\cite{ort15,ort16}. Moreover, the estimated phase of the detected
beatnote is directly linked to the direction of the optical anisotropy
(here, absorption anisotropy) which is responsible for the OB
phenomenon \cite{ort15,ort16}.

The main interest of such OB polarimetric modality is that the
polarimetric information is retrieved from a single acquisition/scan
of the image, hence ensuring easy and fast implementation on existing
imaging setups such as microscopes. However, the polarimetric
information is retrieved from the analysis of a RF-modulated optical
signal, which requires fast PDs and dedicated demodulation electronics
(lock-in detection or quadrature demodulation board) to measure the
in-phase (I) and in-quadrature (Q) signal components. The OBC and
phase informations can finally be retrieved using the following
relations:
\begin{equation}\label{eq:OBCPhi}
  OBC=\frac{\sqrt{I^2+Q^2}}{DC}, \qquad \text{and,}\quad \varphi = \mathrm{atan}\frac{Q}{I}.
  \end{equation}

  The requirement of high-speed detection/demodulation hence hinders
  the use of wide-field cameras, and the technique so far has been
  implemented in a laser scanning imaging configuration, even for
  remote measurement applications \cite{par17}. For this reason, OB
  polarimetric sensing is well suited to be implemented on a confocal
  laser scanning microscope setup, as will be shown in the following.

\subsubsection*{Dichroism/birefringence polarimetric imaging with induced-OB modalities}\label{sec:induced}

The standard OB imaging modality described above has a strong
specificity since OBC contrast can only appear if the sample shows
absorption anisotropy (diattenuation). This can be interesting for a
number of applications. However, in biology where the samples of
interest are rather transparent, dichroism is not the most likely to
occur among other polarimetric effects. To broaden the scope of
application of this unconventional approach, it has recently been
shown that it is possible to gain sensitivity on other polarimetric
effects, such as pure depolarization and birefringence, by slightly
modifying the detection setup \cite{par20}.

These complementary modalities have been referred to as ``induced'' OB
as the RF beatnote carrying the polarimetric information is generated
on an analyzing element placed ahead of the detector, after light has
interacted with the sample. Two interesting modalities have been
identified. The first one, called linear-induced OB (LI-OB), consists
in using a circular left/right DFDP illumination and a linear
polarizer in front of the detector, and allows depolarization
contrasts to be revealed on purely depolarizing samples
\cite{par20}. The second modality, referred to as circular-induced OB
(CI-OB) uses also a circular DFDP illumination, but requires a
circular analyzer in front of the detector. This is commonly obtained
by combining a quarter-wave-plate (QWP) and a linear polarizer with
eigenaxes oriented at 45$^\circ$ from each other. This last modality
has the strong potential to reveal interesting OBC contrast, not only
on dichroic samples, but also on birefringent samples \cite{par20}. In
the remainder of this article, these three OB modalities will be used
and compared in terms of efficiency for polarimetric imaging of cell
samples.

\subsection*{Microscopy setup description}\label{sec:materials}

\begin{figure}
\begin{center}
\begin{tabular}{c}
  \includegraphics[width=14cm]{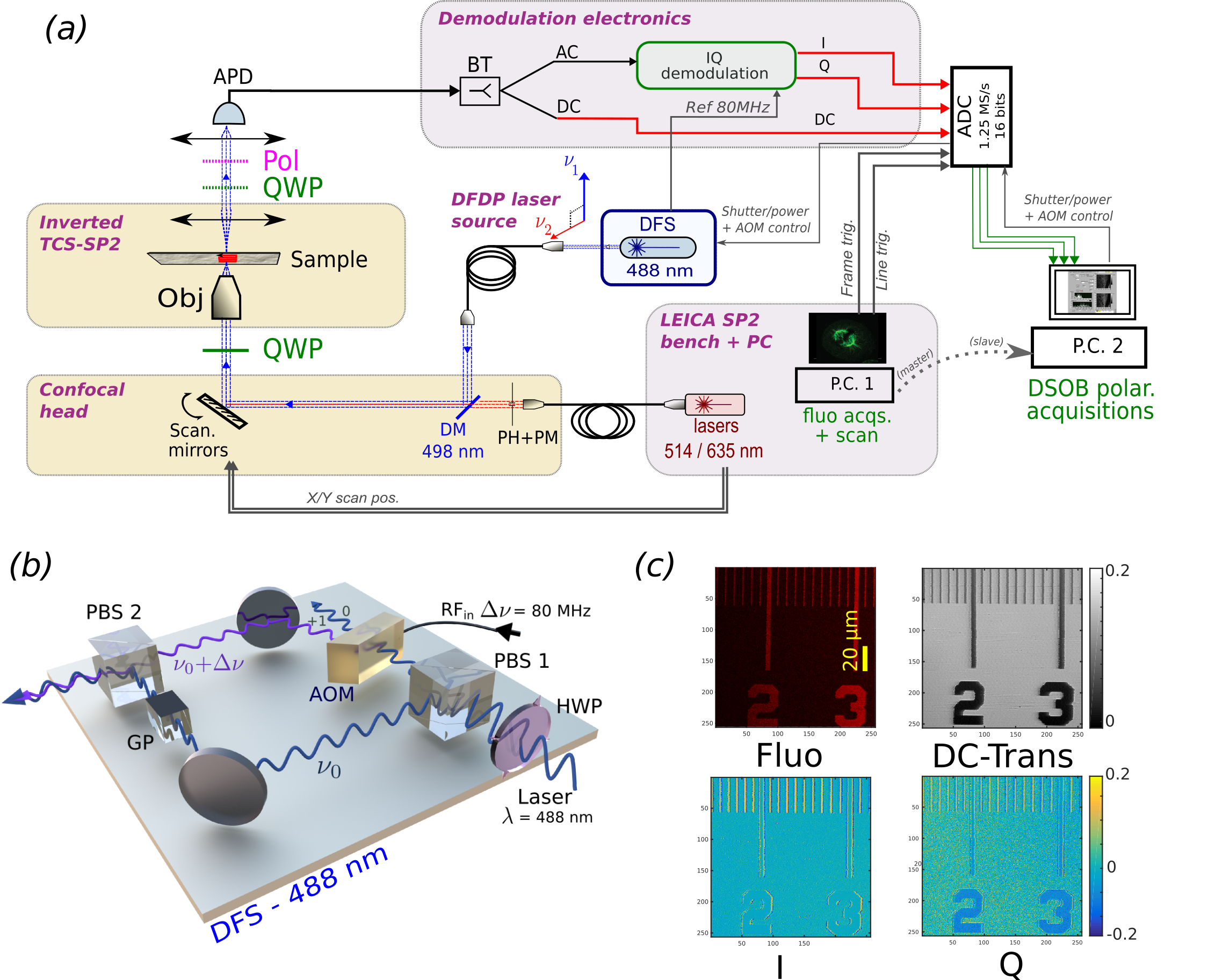}
\end{tabular}
\end{center}
\caption { \label{fig:setup} (a) Sketch of the OB polarimetric
  microscope setup implemented on a standard fluorescence scanning
  microscope (Pol: polarizer, QWP: quarter-wave plate at 488~nm, APD:
  avalanche photodiode, BT: bias-tee, DFS: dual-frequency source, DM:
  dichroic mirror, PH: pinhole, PM: photomultiplier). (b) 3D-sketch of
  the dual-frequency dual-polarization laser source at $\lambda$=488~
  nm designed for the experiment (HWP: half-wave plate, PBS:
  polarization beam splitter, AOM: acousto-optic modulator, GP: Glan
  polarizer). (c) Example of test images recorded from the
  fluorescence confocal microsope (Fluo), and from the polarimetric
  acquisition setup: the DC image provides the transillumination
  image, the I and Q channels are further processed to build the OBC
  and phase images displayed in next figures.}
\end{figure}

In this section, we describe how a standard laser scanning microscope
(Leica TCS-SP2 inverted microscope setup) has been modified in order
to handle OB polarimetric imaging in transmission, while maintaining
the ability to perform confocal fluorescence imaging. This constraint
was necessary to be able to overlay OB and fluorescence images and
thus identify the cellular structures giving rise to the observed
polarimetric contrasts. A sketch of the whole setup is given in
Fig.~\ref{fig:setup}.a.

\subsubsection*{Dual-frequency dual-polarization  illumination}

To perform polarimetric cell imaging in the visible range, we use a
polarization-sensitive Mach-Zehnder based free-space architecture
comprising an acousto-optic shift of 80~MHz in one of the two arms
(acousto-optic modulator, MT80-A1-VIS, AA OPTOELECTRONICS). This setup
is represented on the 3-D sketch of Fig.~\ref{fig:setup}.b allows us
to obtain a linearly-polarized DFDP beam from a 40~mW commercial blue
laser (PC14584, NEWPORT, $\lambda=488$~nm). The DFDP source has previously been
extensively described in \cite{sch14}.

In order to control the optical power deposited on the sample, a
mechanically controlled optical valve has been designed using the
association of a half-wave plate (HWP) (WPHSM05-488, THORLABS) and a
Glan polarizer inserted between the laser output and the
Mach-Zehnder-like setup. The orientation of the Glan polarizer is
adjusted to ensure that the intensity of the two components of the
DFDP beam share equal intensities. The direction of the HWP is
controlled by a rotating mount (URS50 + SMC100 controller, NEWPORT),
allowing remote tuning of the optical power from the LabView (NATIONAL
INSTRUMENTS) program used to control the whole setup.

In order to convey the DFDP beam into the microscope scanning head, we
use a single mode polarization-maintaining optical fiber
(P3-488PM-FC-2, THORLABS) whose eigenaxes have been aligned with the
polarization directions of the DFDP beam. A mechanical shutter
controlled by the LabView software is placed ahead of the fiber
injection and avoids illumination of the sample before an acquisition
is launched, so as to reduce fluorescence bleaching.

\subsubsection*{Microscope setup}

The DFDP beam conveyed by the fiber is coupled into the confocal
microscope (TCS-SP2, LEICA) through the original infrared port using a
dichroic mirror (ZT488rdc, CHROMA, cut-off wavelength 498 nm), as
sketched in Fig.~\ref{fig:setup}.a. This makes it possible to perform
fluorescence and polarimetric imaging simultaneously, as the
excitation laser lines available in the confocal head (514 nm and 635
nm) can be used to illuminate the sample and the fluorescence emitted in
the backward direction can be detected through the confocal
pinhole. The filter wheel of the inverted microscope stand has been
equipped with a removable QWP at 488~nm
(WPQSM05-488, THORLABS), allowing the linear DFDP beam to be converted
into a circular left/right DFDP illumination at the sample
plane. Finally, the light is focused onto the sample by the microscope
objective. The images displayed in the following have been recorded
with one of the two following objectives: a $10 \times$ objective for
test images and polarimetric measurements on synthetic samples
(Fig.~\ref{fig:synthetic}), whereas the images of cells have been
obtained with an oil-immersion $63 \times$ objective.

Upon interaction with the sample, light is detected on the one hand in
the classical confocal backscattering mode for the fluorescence
emission, and the image reconstruction is handled by the TCS-SP2
system. On the other hand, the polarization information is measured in
transmission by detecting the OB beatnote at 80 MHz on an avalanche
photodiode (APD) (Silicium, 400 MHz bandwidth, APD430A, THORLABS), through a
custom optical arrangement comprising the original microscope
condenser lens ($f'$=$28$ mm), a switchable plane mirror, and an
additional focusing lens ($f'$=$17$ mm) (See
Fig.~\ref{fig:setup}.b). This configuration enables full operation of
the LEICA microscope in its original mode, but hinders the ability of
true simultaneous imaging since a slight tuning of the microscope
focus must be ensured when switching between confocal fluorescence
imaging and polarimetric OB imaging.

Finally, in order to implement the three OB modalities described
above, removable polarization analysis elements have been inserted
after the sample, between the condenser lens and the last focusing
lens. A linear polarizer allows LI-OB to be performed, while CI-OB
requires an additional QWP to be inserted before the polarizer, with
eigenaxes oriented at 45$^\circ$ from the polarizer direction.

\subsubsection*{Polarimetric data acquisition and image reconstruction}

As described in the theoretical section, OB polarimetric imaging
requires a specific detection/demodulation chain to recover
the polarimetric information from the estimated amplitude and phase of a
RF beatnote, at 80 MHz in the present case. For that purpose, a first
homemade ``bias-tee'' electronic circuit separates
the 80 MHz AC component of the detect photocurrent from its
continuous-wave DC component. As sketched in Fig.~\ref{fig:setup}.a,
the latter is directly sampled and digitized on the analog-to-digital
conversion (ADC) module of a input/output board (NI-USB 6356, NATIONAL
INSTRUMENTS, 16 bits, 1.25 MS/s per channel).

As for the AC component, it can be demodulated either with a lock-in
amplifier \cite{sch14}, or using a custom-made I/Q demodulation
circuit at the dedicated 80 MHz frequency. Here, we use the I/Q
demodulation approach which has been extensively described in previous
work \cite{par17}. On our microscope setup, the RF reference (local oscillator) used
to demodulated the AC photocurrent is given by the 80 MHz RF signal
that drives the acousto-optic modulator, with appropriate
amplification to match the nominal operating point of the RF mixers
(+7 dBm) involved in the demodulation circuit. At the output of the
I/Q demodulation circuit, the two quadratures (I and Q) are low-pass
filtered (1~MHz cutoff frequency) and further amplified using two
identical switchable gain voltage amplifiers (from $\times 3$ to
$\times 34$) so as to optimally match the input range of the ADC board
which eventually samples and digitizes the I and Q signals along with
the DC component.

The three digitized signals are then processed with a LabView programm
on a computer to build the three raw polarimetric images (DC, I, Q),
from which the OBC amplitude and phase maps are computed using the
relations of Eq.~(\ref{eq:OBCPhi}), taking into account the relative
gain factors and input ranges of the three channels. Data acquisition
and reconstruction of the images are triggered by the ``frame'' and
``line'' trigger signals from the LEICA bench which are also digitized
on the ADC board. In addition, to avoid photodamage or bleaching of
the sample and to minimize aging of the AOM, a $+5$V signal is output
from the I/O board only when the acquisition is started, in order to
open the mechanical shutter of the DFDP source and to enable the RF
high-voltage supply of the AOM.

Using a 200~Hz scanning speed on the LEICA system allows polarimetric
images of $256\times 256$ pixels to be recorded within 1.5~s. In order
to cope with the acceleration/deceleration phases of the galvanometric
mirrors and avoid spatial distortion in the obtained images, the
simplest way consists in acquiring an image format of $512 \times 256$
pixels and remove by software the $128$ first and last pixels from
each line, ending up with distorsion-free $256\times 256$ pixels
images.  An example of raw DC, I and Q images recorded with a
$\times 10$ objective on a stage micrometer is displayed in
Fig.~\ref{fig:setup}.c, along with the corresponding fluorescence
image.

For the sake of accuracy of the estimated OBC amplitude and phase maps, the I
and Q images displayed in the following figures and used to compute
the OBC amplitude and phase values have been corrected by software from slow
phase drifts due to unwanted optical path changes in the DFDP
source. For that purpose, we estimate the slow drift on the quadrature
images by evaluating the local average values of the I and Q signals at four
locations around the cell (typically, the four image corners), and by
using a linear regression to calculate the slow linear trend of the
I/Q signal across the image. This slow trend is finally removed to
provide the I and Q images as those displayed in
Fig.~\ref{fig:compar}.c. Such post-processing steps could be avoided
in future implementation of the system by removing the unwanted
optical phase drifts in the system.

\subsection*{Sample cells preparation and fluorescence imaging}

HeLa and U2OS cells were cultured in Dulbecco’s Modified Eagle’s
Medium (DMEM) supplemented with 10\% fetal bovine serum (FBS),
100~$\mu$g/mL penicillin and 100~U/mL streptomycin and maintainted at
37$^\circ$C in a 5\% CO$_2$ incubator. Cells were seeded on coverslips
24~h prior to fixation. For DRAQ5 staining (U2OS cells), media was
removed and cells were washed once with phosphate buffered saline
(PBS) for 3 min at room temperature. Cells were fixed in 4\%
paraformaldehyde (PFA) for 15 min at room temperature and washed twice
with PBS. Cells were stained with DRAQ5 (0.5 mM in PBS) for 15 minutes
and washed three times with PBS before coverslips were mounted on
slides using ProLong\textregistered\ Gold (THERMO FISHER SCIENTIFIC).
For microtubule staining (HeLa cells), media was removed and cells
were washed once with PBS for 3 min, fixed with 4\% PFA for 15 min and
washed twice with PBS, all performed at room temperature. Cells were
permeablised with 0.2\% Triton X-100 in PBS for 5 min, washed twice
with PBS and placed in blocking buffer (5\% BSA, 0.05\% Tween-20 in
PBS) for 60 min at room temperature. Cells were then incubated with
anti-tubulin antibody (0.2 $\mu$g/mL, sc \#62204, INVITROGEN) in
blocking buffer overnight at 4$^\circ$C. Cells were washed three times
with 0.1\% Triton X-100 in PBS before incubation with Alexa Fluor 488
anti-mouse IgG (2 $\mu$g/mL, A11001, INVITROGEN) diluted in blocking
buffer at room temperature for 1 h in the dark. Cells were washed
twice with 0.1\% Triton X-100 in PBS and counterstained with Hoechst
(1 $\mu$g/mL in PBS) for 10 minutes. Cells were washed three times
with PBS before coverslips were mounted on slides using
ProLong\textregistered\ Gold (Thermo Fisher Scientific).  Fluorescence
excitation of Alexa Fluor 488 and DRAQ5 were excited using 514~nm and
633~nm lasers respectively. For fluorescence detection, we used
bandpass filters adapted to the fluorophore emission spectra.

\section*{Results and Discussion}\label{sec:results}

\begin{figure}
\begin{center}
\begin{tabular}{c}
\includegraphics[width=12cm]{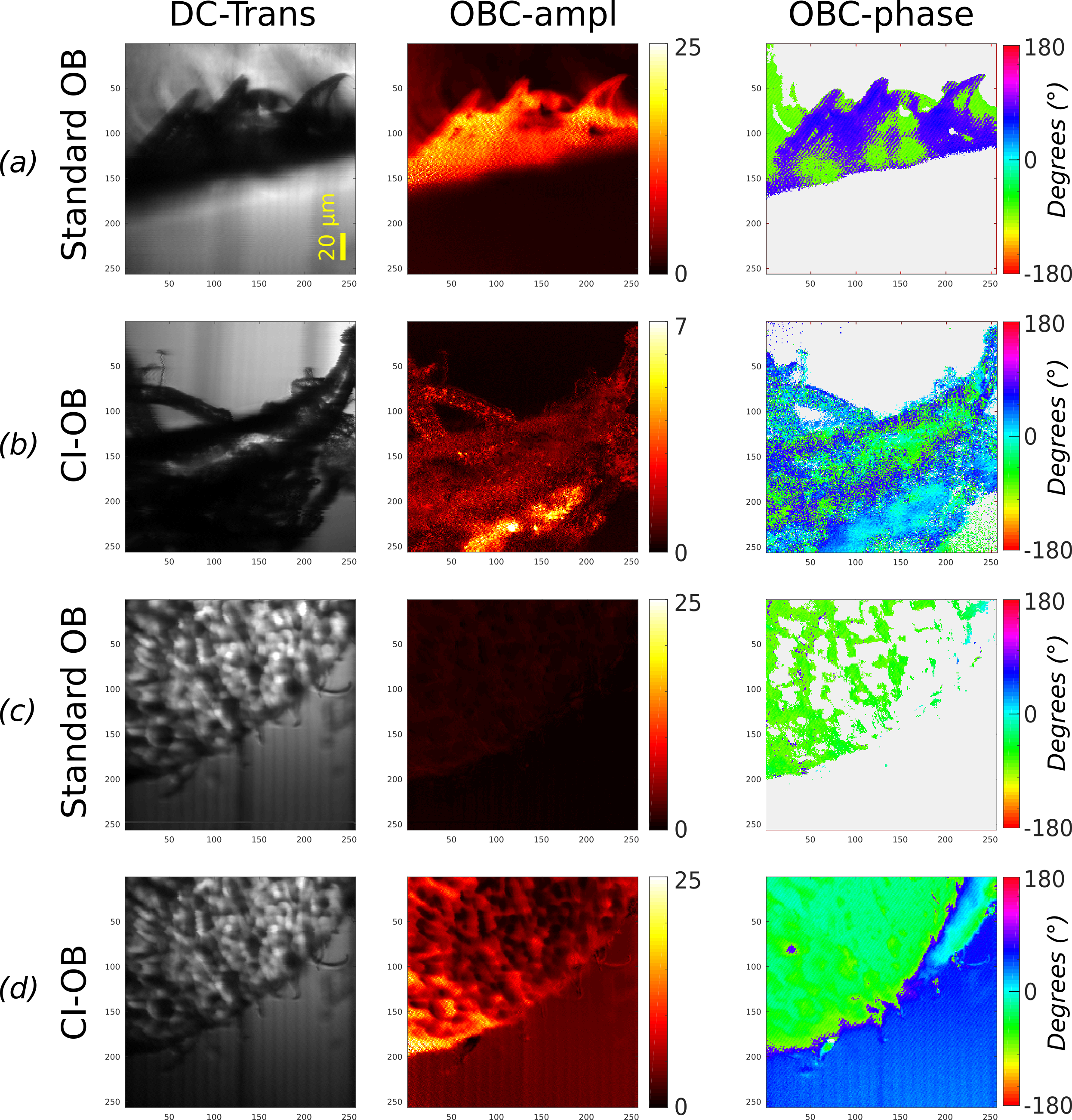}
\end{tabular}
\end{center}
\caption { \label{fig:synthetic} Examples of OB polarimetric images
  recorded with a $10\times$ objective on synthethic samples. Left
  column: DC-transillumination image; Center: OBC amplitude map;
  Right: OBC phase map. (a): standard OB modality on the edge of a
  dichroic sample (tears of polaroid sheets); (b): Circular-induced OB
  modality on a similar sample to (a); (c): standard OB modality on
  cuts of a birefringent sample (plastic tape); (d): Circular-induced
  OB modality on the same sample as in (c).}
\end{figure}

\subsection*{Validation on synthetic samples}\label{sec:validation} 

Before applying the OB polarimetric imaging to cell samples, we
validated the method and the setup on synthetic test samples. These
samples were selected for their known anisotropic optical behaviour,
such as dichroism and birefringence, and the obtained images displayed
in Fig.~\ref{fig:synthetic} are in good agreement with the theory
presented above and in Ref. \cite{par20}. In this figure, the left
column shows the transillumination image obtained with a $\times 10$
objective from the measured DC signal of the APD, while the center and
right column respectively display the OBC amplitude and phase maps obtained from
Eq.~(\ref{eq:OBCPhi}). The phase value being irrelevant when the AC
amplitude is very low (i.e., low OBC), the phase map has been
thresholded and the irrelevant phase values have been represented in
gray color.

The images in the first two rows have been obtained on a dichroic
sample, namely, the cut edges of a polaroid sheet deposited on the
stage micrometer surface, and exhibiting a rather shredded
structure. The standard OB imaging modality has been used in
Fig.~\ref{fig:synthetic}.a, whereas images of
Fig.~\ref{fig:synthetic}.b have been obtained with the
circular-induced modality (CI-OB). As expected, the in-focus parts of
the dichroic sample show strong OB contrast with the two OB
modalities. Theoretically, the estimated phase provides an indication
about the relative orientation of the absorption anisotropies in the
observed sample. The phase maps obtained here showing different values
on the polarizer shreds are in fair agreement with the expected
behaviour.

The second synthetic sample tested corresponds to a piece of plastic
(birefringent) tape sticked on the stage micrometer. The piece of tape
is clearly visible in the upper-left corner of the transillumination
image of Figs.~\ref{fig:synthetic}.c and \ref{fig:synthetic}.d. It has
been imaged here with the standard OB (Fig.~\ref{fig:synthetic}.c) and
the CI-OB modalities (Fig.~\ref{fig:synthetic}.d). Again, as expected
based on the theory, such a birefringent sample does not break the
polarimetric orthogonality, hence resulting in a very low OBC amplitude in
standard OB. However, the birefringent nature of the sample clearly
appears through the significant OBC amplitude obtained on the
birefringent sample with the CI-OB modality. As for the phase maps,
the constant value of the phase estimated on the piece of tape is in
good agreement with the expected uniform orientation of the phase
anisotropy of such sample.

These first imaging results on test synthetic samples thus validate
the correct operation of the polarimetric OB microscope. It can be
noted however here that the amplitude values of OBC (sometimes above
$100 \%$) lack quantitative precision. This would require added
complexity of the system calibration and data processing
\cite{par20,sta20}, but does not prevent qualitative evaluation of the
imaging data gathered on biological samples and presented below.

\begin{figure}
\begin{center}
\begin{tabular}{c}
\includegraphics[width=12cm]{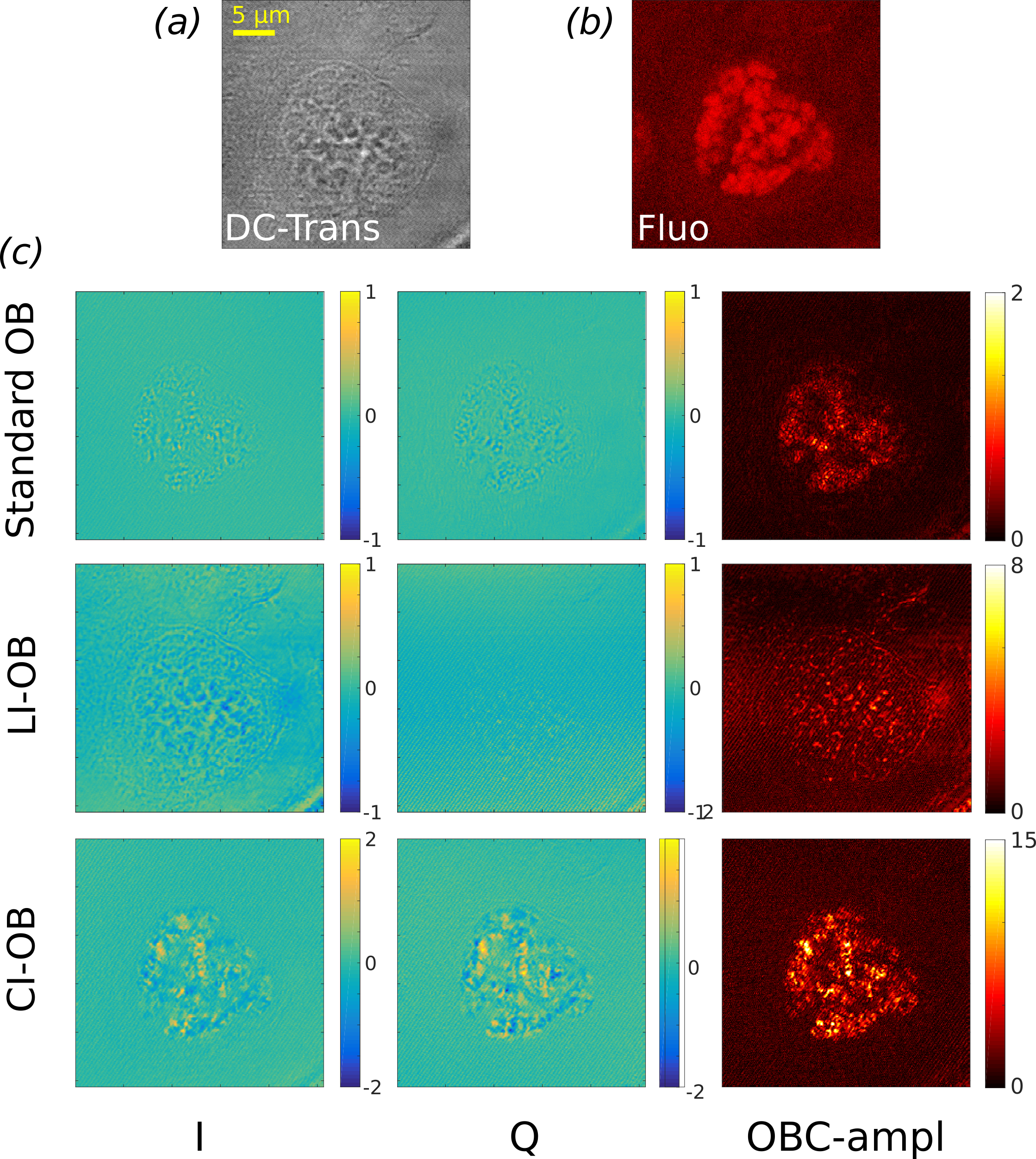}
\end{tabular}
\end{center}
\caption { \label{fig:compar} Fluorescence and OB polarimetric
  images of a U2OS cell whose DNA has been labeled with DRAQ5. Images were obtained
  with a $63\times$ oil-immersion objective. (a) Transillumination image; (b)
  Fluorescence image; (c) OB polarimetric images with I/Q channels and
  OBC amplitude maps for standard OB modality (upper row); linear-induced OB
  (middle row) and circular-induced OB (lower row).}
\end{figure}

\subsection*{OB polarimetric observations in cells}

We imaged human osteosarcoma U2OS cells using the confocal
fluorescence modality of the TCS-SP2 Leica system, and compared the
fluorescence images with the transillumination (DC) and polarimetric
images (OBC amplitude and phase maps) acquired using the method and system
described above. An interesting outcome of these experiments is the
observation of an intrinsic polarimetric OB contrast in mitotic
cells. An example of such acquisition is displayed in
Fig.~\ref{fig:compar}, where a U2OS cell, stained with DRAQ5 for DNA
fluorescence labeling, has been imaged with the confocal fluorescence
microscope (Fig.~\ref{fig:compar}.b), and with the three OB modalities
(Fig.~\ref{fig:compar}.c).  The spatial distribution of the
chromosomes within this cell as seen on the fluorescence image
suggests that it is at an early mitotic stage, most probably
prometaphase.  The analysis of the raw polarimetric images (I and Q)
shows that a moderate OB contrast can be observed in standard OB
imaging modality (first raw of Fig.~\ref{fig:compar}.c). Such OB
contrast seems to arise at compacted chromatin, as shown by the fact
that the spatial distribution of the OBC signal (right column in
Fig.~\ref{fig:compar}.c) resembles the fluorescence image. The
linear-induced OB modality (LI-OB) is inefficient to reveal an
interesting constrast in this structure (middle row), as could be
expected from such a thin and transparent sample for which light
depolarization induced by multiple scattering events during light
propagation is unlikely. In contrast, the circular-induced modality
(CI-OB), which is able to reveal contrast on birefringent samples,
displays the strongest I and Q quadrature signal contributions, and
makes it possible to retrieve a clear OBC amplitude map (lower row, right) with
high signal-to-noise ratio. For this reason, we shall restrict
ourselves in the following to this CI-OB modality, which seems best
adapted to the imaging of compacted chromatin in mitotic cells.

\begin{figure}
\begin{center}
\begin{tabular}{c}
\includegraphics[width=13cm]{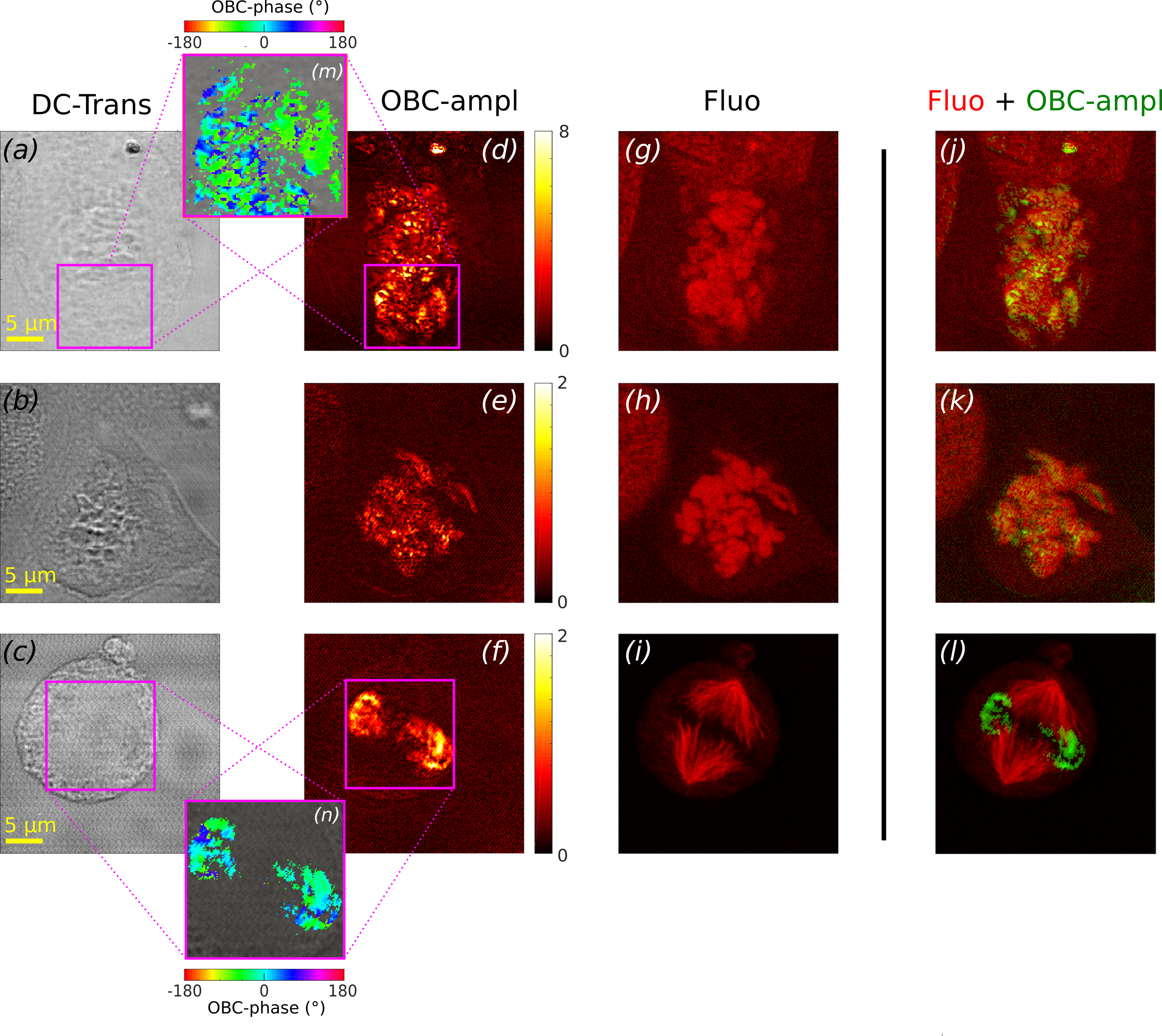}
\end{tabular}
\end{center}
\caption { \label{fig:coloc} Comparison of transillumination (a-b),
  CI-OBC images (c-d) and fluorescence (e-f) acquired on U2OS cells
  labeled with DRAQ5 to highlight the DNA in fluorescence (upper row),
  and on HeLa cells with microtubules labeled by immunofluorescence
  (lower row). In each case, the superimposition of the fluorescence
  and OBC amplitude images (g-h) shows clear colocalization between
  the OBC contrast and the compacted chromatin in mitotic cells. The
  two insets (i-j) show two examples of phase maps extracted from the
  OB images revealing additional morphological contrasts on the
  compacted chromatin.}
\end{figure} 

\subsection*{Label-free polarimetric contrast of mitotic chromosomes}

The clear CI-OB contrast obtained in a dividing U2OS cell such as presented in
Fig.~\ref{fig:compar} is very encouraging towards the possibility to identify mitotic
chromosomes from an endogenous polarimetric contrast, without the need for fluorescence 
labeling. It was however important to verify that such contrast was not specific
to this cell line and not due to the DRAQ5 labeling of the DNA. Moreover, a more
thorough analysis was required to confirm that the polarimetric
contrast colocalizes with the mitotic chromosomes.

For this purpose, we performed some additional acquisitions in two
types of cells: U2OS cells, described in the previous section, and
HeLa cells, a widely used human cancer cell line derived from 
cervix. These cells were labeled either with DRAQ5, to highlight the
DNA (U2OS cells), or with an anti-tubulin antibody, to display the
microtubule network (HeLa cells). We imaged cells undergoing mitosis
in both OBC and fluorescence modalities and the two images were
overlapped to be able to assess the colocalization between the two
signals. The results are displayed in Fig.~\ref{fig:coloc}. It can be
readily observed in Fig.~\ref{fig:coloc}.g that the OBC amplitude map
overlaps very well with the strong fluorescence signal emitted by the
DRAQ5 dye labeling the chromatin. When imaging a dividing HeLa cell
which microtubules were tagged by immunofluorescence, we observed in
Fig.~\ref{fig:coloc}.g the overlay between the OBC amplitude (in green) and the
fluorescence (in red) signals, a characteristic metaphase spindle
composed of microtubules (seen in fluorescence) handling mitotics
chromomes (seen in the polarimetric channel). This last result demonstrates
that the polarimetric contrast observed for the mitotic chromosomes is
not specific to a single cell type and is not due to DNA labeling by
DRAQ5, hence paving the way to label-free imaging of mitotic
chromosomes in dividing cells.

In addition to the OBC maps whose amplitude seems to be related to
chromatin compaction, the OB polarimetric techniques provide
additional information by analyzing the estimated phase of the OB
signal at each pixel of the image \cite{par20}. Two examples of phase
maps (estimated only on pixels showing significant OBC amplitude) are displayed
for a U2OS cell (Fig.~\ref{fig:coloc}.i) and a HeLa cell
(Fig.~\ref{fig:coloc}.j). Interestingly, these two phase images reveal
additional morphological information in the compacted chromatin
that can be observed neither in the OBC amplitude maps nor in the fluorescence
images. Further investigation is required to possibly relate such
observations with morphological/organizational structures in compacted
chromosomes \cite{kir04}.

\begin{figure}
\begin{center}
\begin{tabular}{c}
\includegraphics[width=14cm]{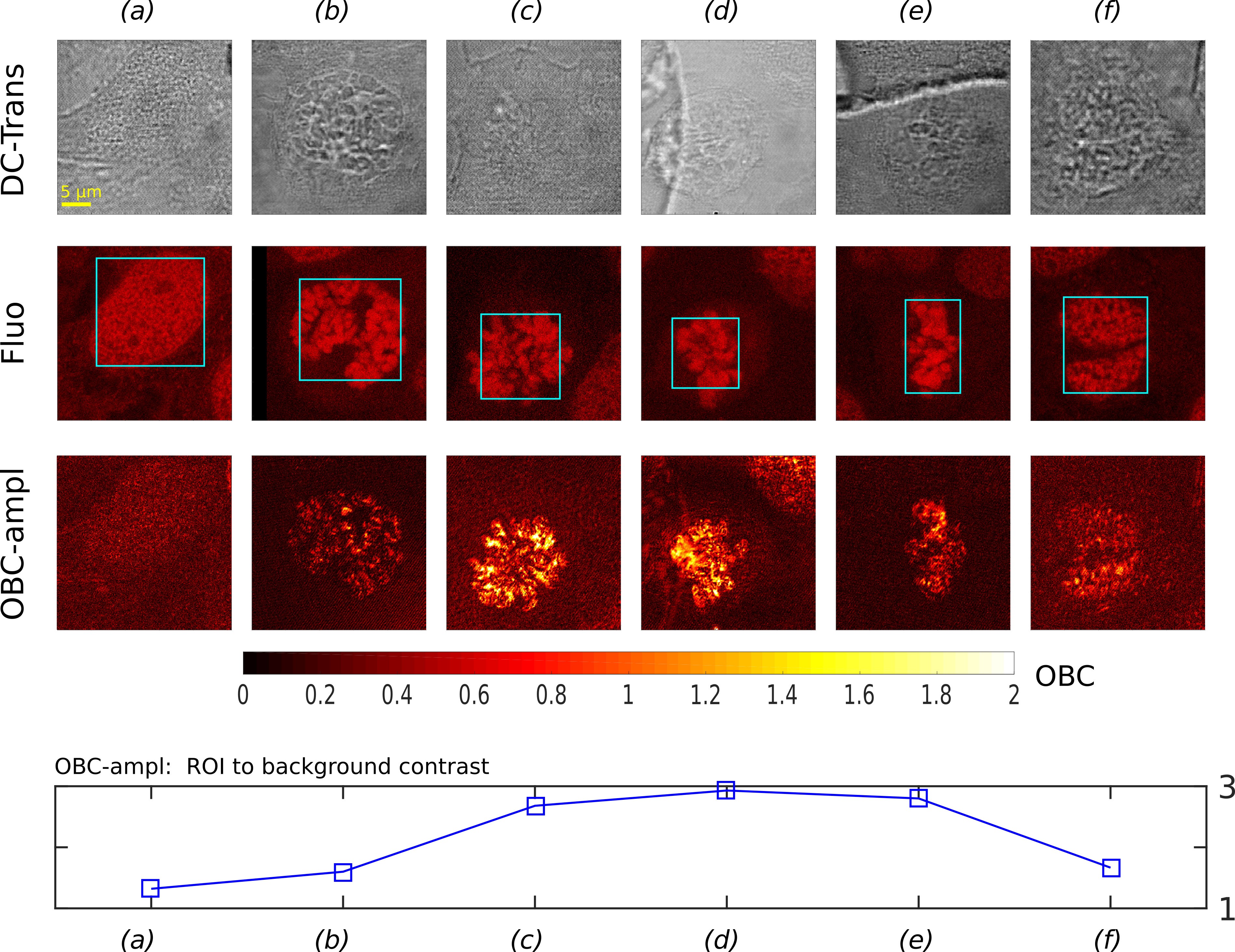}
\end{tabular}
\end{center}
\caption { \label{fig:mitosis} Transillumination, fluorescence and OBC
  amplitude (circular-induced OB) images of DRAQ5 labeled U2OS cells
  at different cell stages: (a) interphase; (b) late prophase; (c)
  prometaphase; (d) late prometaphase; (e) metaphase and (f) late
  telophase. Bottom plot: evaluation of the ROI-to-background contrast
  of the averaged OBC amplitude on the regions displaying DNA in the
  fluorescence images (blue ROIs).}
\end{figure}

As a last experiment to confirm the interest of such endogenous
polarimetric contrast in condensed chromatin, we provide in
Fig.~\ref{fig:mitosis} a comparison of the transillumination (upper
row), OBC amplitude (middle row) and fluorescence (lower row) images
of U2OS cells at different cell stages. From left to right, the
analysis of fluorescence images allows us to identify cells in
interphase (a), late prophase (b), prometaphase (c), late prometaphase
(d), metaphase (e), and finally late telophase (f).  This figure shows
that the polarimetric contrast seems to only arize at compacted
chromatin in cells undergoing mitosis, while no clear OB contrast
could be observed in interphase cells (Fig.~\ref{fig:mitosis}.a). On a
more quantitative basis, the magnitude of the OB polarimetric contrast
seems to follow the chromatin compaction during the mitotic process,
suggesting that OB imaging could be used to monitor the chromatin
compaction state. This is confirmed by the graph displayed at the
bottom of Fig.~\ref{fig:mitosis}. We plotted the ratio of the OBC
amplitude averaged over the regions of interest (ROIs) highlighted in
blue, by its mean value in the surrounding background. The selected
ROIs correspond to the regions where labeled DNA can be identified in
the fluorescence images. With this definition of a ``contrast'' ratio,
the minimum value of 1 corresponds to the absence of any contrast with
respect to the background.

\section*{Conclusion}\label{sec:conclu}

In this article, we have shown that polarized microscopy using
``orthogonality breaking'' approaches could provide valuable
label-free information in biological samples. More specifically, it
has been shown that an endogenous polarimetric
``circular-induced'' OB contrast could be clearly obtained at mitotic
chromosomes during cell division, which has been confirmed by
colocalization with fluorescence images recorded on the same
samples. As the polarimetric modality used in this study is very fast, only
requiring a single scan of the sample, this technique has the potential
to allow real-time live cell imaging to monitor chromosome dynamics
during mitosis. In this context, modifying the microscope setup to
enable imaging in a reflection configuration would ensure strict
simultaneous fluorescence and polarization live-cell imaging.

Applying this label-free technique to study the chromatin compaction
state could open promising perspective for histology studies, such as
the identification of abnormal chromatin compaction arising in some
cancers cells \cite{mea16,rad17}. Further investigation is being
conducted to assess the interest of this approach for imaging other
cellular samples such as embryos \cite{koe11,koi15}, or biological
tissues. This next experimental work will also address the thorough and
rigorous calibration of the setup in order to provide more reliable
quantitative assessments of the estimated OBC amplitude and phase values, which
can be of great interest to investigate the interesting morphological
structures revealed by these two complementary contrasts.



\section*{Authors contributions}

M.T., M.A., S.H. and J.F. designed the research. M.A. and
J.F. designed the laser source as well as the I/Q demodulation RF-setup. M.T., S.D., E.G., G.L.M. and
J.F. designed and assembled the microscope setup. G.L.M., E.G.,
P.R. and J.F. designed the experiment control \& data acquisition
program. R.D., R.S., G.L.M. and S.H. prepared the cell cultures and
samples. R.D., R.S., G.L.M. and J.F. performed the
experiments. J.F. and S.H. performed the data analysis and
interpretation. J.F. and S.H. wrote the paper. All the authors revised
the manuscript.

\section* {Acknowledgments}
The authors would like to acknowledge N.~Ortega-Quijano, F.~Parnet for
their early contribution in assessing the CI-OB modality, and to thank
L.~Frein, S.~Bouhier, C.~Hamel, and A.~Carr\'{e} for their technical
help with the electronical and mechanical design of the
experiments. The authors thank C. Chapuis for her technical assistance
in sample preparation and X.~Pinson for his technical help with the
confocal microscope setup. The authors thank the MRic facility from
the BIOSIT joint unit of services, the University of Rennes 1 and the
CNRS \emph{“Mission pour les Initiatives Transverses et
  Interdisciplinaires”} (MITI) for funding this project. R.D. was
supported by the Universit\'{e} Bretagne Loire, the R\'{e}gion
Bretagne and the Institut Universitaire de France. R.S. was supported
by the PRESTIGE program coordinated by Campus France
[PRESTIGE-2017-2-0042], the Universit\'{e} Bretagne Loire and the
Fondation ARC pour la recherche sur le cancer [PDF20181208405]. We
thank the GDR Imabio for funding P.R.'s and E.G.'s internships. The
MRic facility is member of the national infrastructure
France-BioImaging supported by the French National Research Agency
(ANR-10-INSB-04).




\end{document}